# Free Energy of ATP hydrolysis manipulates the cellular calcium signals

Yingda Ge, Congjian Ni, Yunsheng Sun, Fangting Li[1]

**ABSTRACT** In living cells, oscillation of the concentration of cytosolic $Ca^{2+}$ is an important and pervasive signal for the intercellular and intracellular information conduction. To generate the oscillation, the hydrolysis of ATP is always needed. Many recent studies show that both ATP molecules themselves and the free energy by ATP hydrolysis play significant role in the biochemical process involving ATP hydrolysis. To verify the prediction, we consider the role of ATP molecules and their hydrolysis in a classic one-pool model of $Ca^{2+}$ oscillation. Our results show that the available Gibbs free energy of ATP hydrolysis (ΔG), which measures the "distance" of a reaction to its equilibrium state, is another important regulatory factor of the oscillation system besides the concentration of ATP. Furthermore, our model suggest a rudimental prediction of how the oscillation system changes in an aging cell, such as the decrease of the amplitude and the increase of the least ΔG required for the oscillation.

# 1. Introduction

Hydrolysis of ATP (Adenosine Triphosphate) is the major energy source in living cells. Besides supplying energy to basic activities of cells, however, ATP hydrolysis can modulate intracellular signals by changing the level of the available Gibbs free energy (ΔG) of the reaction of ATP, ADP (adenosine diphosphate) and phosphate. Precisely, in many cases, it is not just the concentration of ATP molecules that modulates a biochemical process, but the ΔG of the whole hydrolysis reaction that does. Taking PdPC (phosphorylation-dephosphorylation cycle) as an example, its ultrasensitivity as a biochemical switch is directly controlled by ΔG rather than the concentration of ATP molecule [1].

Recently, more and more studies show that ΔG is an important regulatory factor in biochemical process. $Ca^{2+}$ oscillation, a vital signal in living cells for example, is also closely related to ATP and its hydrolysis reaction. In cardiac myocytes, calcium spark and wave are closely related to the contraction and relaxation [2]; in islet cells, calcium oscillation regulates the release of many kinds of hormone [3]; and calcium signal is also related to the excitation of neurons [4]. Although few researchers focus on the relationship between $Ca^{2+}$ oscillation and the available free energy of ATP hydrolysis, the amount of energy consumption is actually an important factor controlling the oscillatory system. Analysis on it can help us draw a clearer pattern of the modulation of oscillation, moreover, the whole intercellular and intracellular information pathway.

[1] School of Physics, Center for Quantitative Biology, Peking University, Beijing 100871, China

In this article, we improve a basic model of $Ca^{2+}$ oscillation [5] by adding details of ATP and its hydrolysis. And then we use computer simulation combined with some former study results to discuss the effect of ΔG of ATP hydrolysis to the modulation of intracellular $Ca^{2+}$ oscillation. According to our simulation result, ΔG and concentration of ATP itself function differently in the modulation of intracellular $Ca^{2+}$ oscillation. This result may provide some new thoughts in the study of $Ca^{2+}$ oscillation in some kinds of cells that are very sensitive to the difference of energy level.

# 2. Model

The structure of the network is shown in Fig.1. When cells are in modest condition, most $Ca^{2+}$ are stored in intracellular calcium-stores, e.g. endoplasmic reticulum (ER). Once a specific kind of agonists near the cell membrane bind to cell membrane receptors, a series of reactions will be elicited near membrane and cause a rise in the concentration of inositol 1,4,5-trisphosphate (InsP3 or InsP3), also recognized as second messenger, carrying the information brought by agonists. There is a kind of $Ca^{2+}$ channels on ER membrane, activated as InsP3 molecules binding on its specific sites. These channels are noted as InsP3 receptor, or InsP3R. InsP3R could be regulated by cytoplasmic $Ca^{2+}$ and adenosine triphosphate (ATP) also by binding on specific sites. Thus there would be a rise in cytoplasmic $Ca^{2+}$ concentration ($[Ca^{2+}]_{cyto}$), then amplified by released $Ca^{2+}$ themselves. As $[Ca^{2+}]_{cyto}$ growing up, SERCAs(sarco/endoplasmic reticulum Ca2+-ATPase), a kind of $Ca^{2+}$ pumps on ER membrane and PMCAs(plasma membrane Ca2+-ATPase), which are on cell membrane (plasma membrane), will transport these ions back to store or out of the cell using the energy supported by ATP hydrolysis. Intuitively thinking, these two process, a positive feedback by InsP3R and a negative feedback by ATPase, could cause oscillations in $[Ca^{2+}]_{cyto}$ under certain conditions.

This model is generalized, which does not indicate any specific kinds of cells, but the structure and network shown above is always the core part of many kinds of models of intracellular calcium oscillation and wave[5]. Moreover, the model follows three basic assumptions: $Ca^{2+}$ concentration in cytoplasm ($[Ca^{2+}]_{cyto}$) and ER ($[Ca^{2+}]_{er}$) are considered as variables, while plasmatic $Ca^{2+}$ ($[Ca^{2+}]_{PM}$)are set as a constant; The time scale of our model is restrict to the range that $Ca^{2+}$ signal's feedback on metabolism could be ignored, i.e. the concentrations of ATP, ADP and phosphate are treated as adjustable parameters rather than variables; The concentration of each species is assumed to be homogeneous, so the transduction of the spatial wave is ignored.

## 2.1 Channels on cell membrane

There are mainly three kinds of Ca2+ channels on cell membrane, SOCC (Store-Operated Ca2+ channels), ROCC (receptor-operated Ca2+ channels) and VGCC (voltage-gated Ca2+ channels). Since membrane potential is not important to what we are concerned about and its effect is relatively independent [6], we ignore membrane potential in our model. Comparing to ROCC, we have less knowledge of the details and importance of SOCC [5]. Based on these reasons, we only include ROCC in our model (Fig. 1).

Since GPCRs (G protein-coupled receptor) on cell membrane regulate both the producing of InsP3 and the opening of the corresponding $Ca^{2+}$ channels [7]. (Dupont and Goldbeter, 1993) We retain their simplest assumption that the influx from the extracellular medium triggered by external stimulation is proportional to InsP3's concentration:

$$I_{in} = V_{in} + k_{in}[InsP3] , \qquad (1)$$

where $V_{in}$ represents the constant Ca2+ influx through ROCCs, and the second term represents the part of Ca2+ influx that has positive correlation with the concentration of InsP3, which is simplified to be proportional.

## 2.2 16-state model of InsP3 receptor

InsP3R channel is a kind of $Ca^{2+}$ channel on ER membrane, which is a tetramer of four InsP3R molecules [8]. Under the manipulation by InsP3, $Ca^{2+}$ and ATP molecules, InsP3R plays an important role in the generation and manipulation of intracellular $Ca^{2+}$ oscillation.

In the classic model, they did not consider how ATP molecules manipulate InsP3R. However, since we have to carefully distinguish the function of ATP molecules and ATP hydrolysis, more details of InsP3R have to be involved in this model. There are several kinds of hypo types of InsP3R, and they have similar response to InsP3, $Ca^{2+}$ and ATP: With the growing of InsP3 concentration, the opening probability of InsP3R channel will increase until InsP3 gets saturate. As for $Ca^{2+}$, the opening probability will be relatively lower when $Ca^{2+}$ concentration is too low or too high, particular values of which varies from different kinds of InsP3R. With the increase of ATP, however, the channel will still open when $Ca^{2+}$ concentration is lower [8].

Based on the facts above and an 8-state model of InsP3R [9] assuming that each InsP3R molecule contains one InsP3 binding site, one $Ca^{2+}$ activating binding site and one $Ca^{2+}$ inhibiting binding site, we develop a 16-state model of InsP3R by adding an ATP binding site. When the ATP binding site is occupied, $Ca^{2+}$ are more likely to bind on the activating binding site, which makes InsP3R open even when $Ca^{2+}$ concentration is relatively lower than usual.

For convenience of description, we abbreviate the occupation state of a InsP3R monomer into a four-bits array in square brackets like $S_{c_1c_2pa},(c_1,c_2,p,a=0,1)$, in which 1 is for occupied state, and 0 for unoccupied state. And $c_1, c_2, p, a$ indicate $Ca^{2+}$ activation site, $Ca^{2+}$ inhibition site, InsP3 binding site, ATP binding site. $\sum_{c_1,c_2,p,a} S_{c_1c_2pa}=1$ Once $Ca^{2+}$ inhibition site $(c_2)$ is occupied, InsP3R channel will always be closed, otherwise the channel's opening possibility will be nonzero, but still regulated by $Ca^{2+}$, InsP3 and ATP. (Fig. 2)

Furthermore, the rate constant of each binding process is in different order of magnitudes [10]. Therefore, quasi-steady assumption suits this situation. After applying quasi-steady assumption to each binding process, the equation can be simplified as (Appendix A):

$$\frac{dh}{dt}=-\phi_1 h+\phi_2(1-h) \qquad (2)$$

$$h=\sum_{c_1,p,a} S_{c_1 0 pa} \qquad (2.1)$$

$$\phi_1=\frac{k_3[Ca^{2+}]+k_4[Ca^{2+}][InsP3]/K_5}{1+[InsP3]/K_5} \qquad (2.2)$$

$$\phi_2=\frac{k_{-3}[Ca^{2+}]+k_{-4}[Ca^{2+}][InsP3]/K_6}{1+[InsP3]/K_6} \qquad (2.3)$$

$$K_i=k_{-i}/k_i, i=1,2,5,6,7. \qquad (2.4)$$

We assume that when monomer states $S_{1000}S_{1010}S_{1001}S_{1011}$ are activated, and only then, the $Ca^{2+}$ could go through InP3R. The open probability and transition intensity of the channel are

$$P_0=\left(h\frac{[Ca^{2+}]/K_1+[Ca^{2+}][ATP]/(K_2K_7)}{1+[Ca^{2+}]/K_1+[ATP]/K_7+[Ca^{2+}][ATP]/(K_2K_7)}\right)^4 \qquad (3)$$

$$I_{InsP3R}=V_{InsP3R}P_o([Ca^{2+}]_{er}-[Ca^{2+}]_{cyto}), \qquad (4)$$

where the intensity is in direct proportional to open probability and concentration gradient between both sides of channel [10].

## 2.3 $Ca^{2+}$ pumps (SERCA/PMCA)

$Ca^{2+}$ pump, which can transport $Ca^{2+}$ through membrane against concentration gradient using the energy provided by ATP hydrolysis, is an important part of the whole oscillation system. As the only part that consumes energy of ATP hydrolysis in the model, $Ca^{2+}$ pump is indispensable and needs detailed analysis.

SERCA and PMCA are two kinds of $Ca^{2+}$-ATPase respectively locate on ER membrane and cell membrane. In a general model of $Ca^{2+}$ oscillation, there is no necessary to distinguish the detailed difference of structure between them [5], so in this model, they share the same chemical equations as (simplified from the study of Inesi, G., 1987 [11]):

$$2Ca^{2+}_{low} + E_1 + ATP \underset{k_{-1}}{\overset{k_1}{\rightleftharpoons}} E_2 + ADP \qquad (5)$$

$$E_2 \underset{k_{-2}}{\overset{k_2}{\rightleftharpoons}} 2Ca^{2+}_{high} + E_1 + Pi\ . \qquad (6)$$

Where $Ca^{2+}_{low}$ indicates the $Ca^{2+}$ concentration where the concentration gradient is relatively lower (i.e. cytoplasm), and $Ca^{2+}_{high}$ indicates the $Ca^{2+}$ concentration where the concentration gradient is relatively higher (i.e. extracellular matrix and endoplasmic reticulum). $E_1$ and $E_2$ indicate different conformations of $Ca^{2+}$-ATPase, where $E_1$ indicates the $Ca^{2+}$-ATPase that is not compounded by $Ca^{2+}$ and phosphate, and in state $E_2$, the $Ca^{2+}$-ATPase is compounded by two $Ca^{2+}$ and one phosphate molecule.

Taking SERCA as an example, $Ca^{2+}_{high}$ and $Ca^{2+}_{low}$ are replaced by $Ca^{2+}_{er}$ and $Ca^{2+}_{cyto}$, which respectively indicate $Ca^{2+}$ concentration in ER and cytoplasm. After a series of calculation (Appendix B), we have:

$$I_{SERCA} = \frac{d[Ca^{2+}]_{er}}{dt} = 2E_T \frac{k_1 k_2 [ATP][Ca^{2+}]^2_{cyto} - k_{-1}k_{-2}[ADP][Pi][Ca^{2+}]^2_{er}}{k_1[ATP][Ca^{2+}]^2_{cyto} + k_{-2}[Pi][Ca^{2+}]^2_{er} + k_{-1}[ADP] + k_2}, \qquad (7)$$

where $E_T$ is proportion to the total number of SERCAs.

In order to investigate how ΔG of ATP hydrolysis affects the $Ca^{2+}$ signals, it is necessary to illustrate the relationship between ΔG and the function of $Ca^{2+}$ ATPase.

From equation above, we have:

$$I_{SERCA} = 2E_T \frac{\frac{k_1 k_2 [ATP][Ca^{2+}]^2_{cyto}}{k_{-1}k_{-2}[ADP][Pi][Ca^{2+}]^2_{er}} - 1}{\frac{k_1[ATP][Ca^{2+}]^2_{cyto}}{k_{-1}k_{-2}[ADP][Pi][Ca^{2+}]^2_{er}} + \frac{1}{k_{-1}[ADP]} + \frac{1}{k_{-2}[Pi][Ca^{2+}]^2_{er}} + \frac{k_2}{k_{-1}k_{-2}[ADP][Pi][Ca^{2+}]^2_{er}}} \quad (8)$$

Let $\frac{k_1 k_2 [ATP][Ca^{2+}]^2_{cyto}}{k_{-1}k_{-2}[ADP][Pi][Ca^{2+}]^2_{er}} = \gamma_0$ , then $\Delta G_{whole} = K_B T \ln \gamma_0$ represents the difference of the whole Gibbs free energy when one ATP molecule is hydrolyzed into an ADP molecule and a phosphate molecule, which includes the energy released by the hydrolysis of ATP and the work transporting $Ca^{2+}$ against concentration gradient.

Moreover, if we define $\frac{k_1 k_2 [ATP]}{k_{-1}k_{-2}[ADP][Pi]} = \gamma$, then the free energy of one molecule ATP hydrolysis can be written as:

$$\Delta G_{whole} = K_B T \ln \gamma_0 = K_B T \ln(\gamma \frac{[Ca^{2+}]^2_{cyto}}{[Ca^{2+}]^2_{er}}) = K_B T \ln \gamma + 2K_B T \ln(\frac{[Ca^{2+}]_{cyto}}{[Ca^{2+}]_{er}}) \triangleq \Delta G + W.$$

The first term represents the energy release of ATP hydrolysis, and the second term represents the work of transporting $Ca^{2+}$ against concentration gradient. Thus we get

$$I_{SERCA} = 2E_T \frac{\gamma \frac{[Ca^{2+}]^2_{cyto}}{[Ca^{2+}]^2_{er}} - 1}{\frac{\gamma [Ca^{2+}]^2_{cyto}}{k_2 [Ca^{2+}]^2_{er}} + \frac{k_{-2}\gamma [Pi]}{k_1 k_2 [ATP]} + \frac{1}{k_{-2}[Pi][Ca^{2+}]^2_{er}} + \frac{\gamma}{k_1 [ATP][Ca^{2+}]^2_{er}}} \quad (9)$$

From the equation above, we can find that the farther ATP hydrolysis is away from equilibrium state, the faster Ca2+ can be transported by Ca2+ pumps.

If we replace Boltzmann constant with universal gas constant, R, then ΔG indicates the difference of free energy in the hydrolysis reaction of 1 mol ATP.

## 2.4 Leaking

Besides the channels and pumps mentioned above, $Ca^{2+}$ can pass through many other channels driven by the gradient of its concentration. But those channels have little direct connection with this oscillation system, thus we assume that the flux is proportional to the concentration gradient in order to simplify our model:

$$I_{leak} = V_{leak}([Ca^{2+}]_{er} - [Ca^{2+}]_{cyto}) \quad (10)$$

## 2.5 Total model

Assembling all parts mentioned above, the system can be described by three variables: $[Ca^{2+}]_{cyto}$, $[Ca^{2+}]_{er}$ and h. Considering the volume difference between cytosol and ER, an ratio parameter $\beta$ is involved to describe the flux of$[Ca^{2+}]_{er}$.

$$\frac{d[Ca^{2+}]_{cyto}}{dt} = -I_{SERCA} - I_{PMCA} + I_{InsP3R} + I_{leak} + I_{in} \tag{11}$$

$$\frac{d[Ca^{2+}]_{er}}{dt} = \beta(I_{SERCA} - I_{InsP3R} - I_{leak}) \tag{12}$$

$$\frac{dh}{dt} = -\phi_1 h + \phi_2(1-h) \tag{13}$$

## 2.6 Parameters

Table 1 shows the values of the parameters used in this model, where AD means that there are few solid resources of its value, and this parameter is adjusted to suit this model. For most of those adjusted parameters, we refer to the book *Models of Calcium Signaling (2016)* [5], and we adjust them to make the channels have familiar response to signals as those in this book do.

.

Table 1 Parameters of this model

| Name | Value(unit) | Reference | Name | Value(unit) | Reference |
|---|---|---|---|---|---|
| $\boldsymbol{K_1}$ | $0.0328(\mu M)$ | [9] | $\boldsymbol{V_{in}}$ | $0.02$ | **AD** |
| $\boldsymbol{K_2}$ | $0.2$ | **[9]** | $K_{in}$ | $0.0004(\mu M)$ | **AD** |
| $\boldsymbol{k_{+3}}$ | $1.5 \times 10^{-2}(\mu M^{-1}s^{-1})$ | [9] | $\boldsymbol{k_1}$ | $1 \times 10^{3}(\mu M^{-3}s^{-1})$ | **[11]** |
| $\boldsymbol{k_{-3}}$ | $2 \times 10^{-3}(s^{-1})$ | [9] | $\boldsymbol{k_{-1}}$ | $6.4 \times 10^{-4}(\mu M^{-1}s^{-1})$ | **[11]** |
| $\boldsymbol{k_{+4}}$ | $1.5 \times 10^{-1}(\mu M^{-1}s^{-1})$ | [9] | $\boldsymbol{k_2}$ | $8 \times 10^{2}(s^{-1})$ | **[11]** |
| $\boldsymbol{k_{-4}}$ | $2 \times 10^{-1}(s^{-1})$ | **[9]** | $\boldsymbol{k}_{-2}$ | $6.25(\mu M^{-3}s^{-1})$ | **[11]** |
| $\boldsymbol{K_5}$ | $0.1(\mu M)$ | **[9]** | $E_{ER_T}$ | $7.5 \times 10^{-3}(\mu M)$ | **AD** |
| $K_6$ | $0.943(\mu M)$ | **[9]** | $E_{PM_T}$ | $5\times10^{-3}(\mu M)$ | **AD** |
| $K_7$ | $0.7$ | **AD** | $[Pi]$ | $1\times10^{3}(\mu M)$ | **[13]** |
| $V_{InsP3R}$ | $1\times10^{2}(s^{-1})$ | **AD** | $K_{eq}$ | $4.9\times10^{11}(\mu M)$ | **[1]** |
| $[InsP3]$ | $0.1(\mu M)$ | **[12]** | $[Ca^{2+}_{PM}]$ | $40(\mu M)$ | **[5]** |
| $V_{leak}$ | $6\times10^{-4}(s^{-1})$ | **AD** | | | |

# 3 Results

## 3.1 Time sequence simulation

While using the parameters in Table 1 and initial values of $[Ca^{2+}_{er}]$=9μM, $[Ca^{2+}_{cyto}]$=0.07μM and h=0.9, we can get an oscillatory simulation result (Fig. 3). Actually, this model is very inertial to the change of initial values, so it is safe to use the same initial values in our simulation. The concentration of ATP in the simulation is 1600μM, and the value of ΔG=23RT=57kJ/mol, where T=300K. They are very close to the values in living cells. To make it clear, the function of agonist is indicated by the concentration of InsP3. In this case, the concentration of $Ca^{2+}$ will oscillates between a high level (~0.5μM) and a lower level (~0.1μM) in a frequency of about 20 seconds. Comparing it with some former study results, the frequency and amplitude both fit well [14]. It means that our model can well simulate Ca2+ oscillation in normal cells when adding the details of the mechanism of Ca2+ pumps and InsP3 receptors to it. This would make our latter results more convictive.

## 3.2 Parameter sensitivity analysis

After analyzing parameter sensitivity of this model (Appendix C), we find that most of our parameters are relatively stable, while there are four sensitive parameters, $k_1$, $k_{-2}$, $K_2$ and $V_{InsP3R}$. These parameters are either about the $Ca^{2+}$ flux InsP3R or $Ca^{2+}$ pumps, and most of them have close relationship with the binding of $Ca^{2+}$, which means that the binding of $Ca^{2+}$ to InsP3R and $Ca^{2+}$ pumps plays an important role in the manipulation of $Ca^{2+}$ oscillation. What is non-negligible is that ATP and its hydrolysis may also markedly influence the function of these two parts. So ATP and ΔG should also be carefully considered while doing parameter sensitivity analysis.

To visualize the influence of those sensitive parameters to the whole system, we choose bifurcation graphs to show how the system changes while adjusting the parameters (Fig.5). According to the analysis of bifurcation behavior of those parameters, we find that there is a Hopf bifurcation near the bifurcation point.

By analyzing the bifurcation graphs, we find that although those four parameters are relatively more sensitive than others, their values are still far from the bifurcation points, as for ATP and ΔG, however, their values are very close to the bifurcation points, which means that a slight change of their value may decide whether the system is in a stable state or an oscillatory state. That's one of the reasons why we focus our eyes on the effect of them.

## 3.3 Analysis of ATP and ΔG

Since ATP and ΔG are important for $Ca^{2+}$ oscillation, detailed analysis of them is needed. In order to analyze the connection between them, we plot more bifurcation graphs about them. While plotting the bifurcation graphs of ΔG, we control the concentration of ATP at 3 different values, and vice versa (Fig. 5). With the growing of ATP concentration, the sensitivity to ΔG of the system increases; and with the growing of ΔG, sensitivity to ATP also increases. This result shows that ATP concentration and ΔG of ATP hydrolysis are both important factors in manipulation of $Ca^{2+}$ oscillation.

Furthermore, to investigate the overall correlation of ATP concentration and value of ΔG, we make ATP and ΔG as two independent parameters (Fig. 6), which could make sense because the concentration of phosphate is always kept saturated in living cells, there are two independent parameters needed to determine the state of ATP hydrolysis.

According to the difference of the factors that mainly manipulate the oscillation, Fig 6 could be divided into three areas. One is ATP controlling area, which is in the upper part of the graph where ΔG is relatively high (more than 24RT). In this area, the bifurcation, amplitude and frequency are all controlled by ATP concentration, while the change of ΔG has little influence to them. Another one is ΔG controlling area, which is in the left part of the figure where ATP concentration is high (more than 3 mM). Here this system is controlled by ΔG. The last part is in the lower left part of the graph, where ATP and ΔG both can significantly control the behavior of the oscillatory system. The values of ATP and ΔG in this part are similar to what in living cells. It indicates that in living cells, ATP concentration and ΔG are both vital manipulators of $Ca^{2+}$ oscillation system. Additionally, there exists a minimum ATP concentration and a minimum ΔG level of the oscillation zone, only when the ATP concentration and ΔG are both higher than this threshold can this system be in an oscillatory state, which means that ATP and ΔG both are important and independent switches of calcium oscillation. Combining to the definition and biophysical meaning of ΔG, this result indicates that how far the ATP hydrolysis reaction is from its equilibrium state determines whether this system can enter an oscillatory state.

We may use eq. 9 to explain the distribution of the three areas. When ΔG is getting higher, meaning the system is getting further from equilibrium state, γ will be considerably larger than other items in eq. 9, so that γ will be eliminated and the whole system would only be manipulated by ATP concentration. The same thing also happens to ATP concentration when it is relatively high. If we check the dependence of the value of $I_{SERCA}$ in eq. 9 to the values of ATP concentration and γ, which is shown in fig. 7A, we can get this conclusion more directly.

Moreover, although it is hard to get the analytic result of the bifurcation behavior in this model, we can at least draw the analytic expression of the fixed points. Setting eq. 11 to 13 to be zero, we can get the concentration of plasmatic calcium to be:

$$[Ca^{2+}]_{cyto} = \sqrt{\frac{[2E_{pm} + \frac{k_{-2}\gamma[Pi]}{k_1 k_2 [ATP]}(V_{in} + k_{in}[InsP3])][Ca^{2+}]_{pm}^2 + (\frac{1}{k_2[Pi]} + \frac{\gamma}{k_1[ATP]})(V_{in} + k_{in}[InsP3])}{(2E_{pm} - \frac{V_{in} + k_{in}[InsP3]}{k_2})\gamma}} \tag{14}$$

When ATP goes towards infinite,

$$[Ca^{2+}]_{cyto} = \sqrt{\frac{2E_{pm}[Ca^{2+}]_{pm}^2 + \frac{1}{k_2[Pi]}(V_{in} + k_{in}[InsP3])}{(2E_{pm} - \frac{V_{in} + k_{in}[InsP3]}{k_2})\gamma}} \propto \gamma^{-1/2} \tag{15}$$

And if γ approaches infinite,

$$[Ca^{2+}]_{cyto} = \sqrt{\frac{\frac{k_{-2}[Pi]}{k_1 k_2 [ATP]}(V_{in} + k_{in}[InsP3])[Ca^{2+}]_{pm}^2 + \frac{1}{k_1[ATP]}(V_{in} + k_{in}[InsP3])}{(2E_{pm} - \frac{V_{in} + k_{in}[InsP3]}{k_2})}} \propto [ATP]^{-1/2} \tag{16}$$

The results above indicate that the position of the fixed points of $Ca^{2+}$ in cytoplasm is directly manipulated by the concentration of ATP and the level of γ. And when one of them is relatively high, it's totally controlled by the other.

When ATP concentration is around 1mM, which is believed to be close to its concentration in living cells, whether the system is in oscillatory state sensitively depends on the value of ΔG. And while the value of ATP concentration is getting higher, ΔG could markedly manipulate the signal of $Ca^{2+}$ oscillation.

Considering the definition of ΔG, we can safely draw the conclusion that the effect of ΔG appears in its manipulation of the function of $Ca^{2+}$ pumps (as for InsP3R, ATP just binds on it rather than hydrolyzes there [15]). To make the effect of ΔG more apparent, we shut down InsP3 receptor by setting h to zero. Then the whole system is mainly regulated by the activity of Ca2+-ATPase (SERCAs and PMCAs). Because of the loss of $I_{InsP3R}$, the important positive feedback of $Ca^{2+}$ signal, oscillation could not be found. So we take the stable solution of $Ca^{2+}$ concentration instead (Fig. 7B).

As the concentration of ATP getting down, there is a bending of the boundary. According to the previous analytic solution of SERCA/PMCA, the bending is caused by the nonzero constant value of leaking rate, no longer the covalent binding of ATP. Therefore, in this situation, the whole system is affected mainly by ΔG, which is proved in both analyzing and computing ways.

### 3.4 Simulation of aging cells

Within the progress of aging, there will be a lot of changes in a cell. Some of the changes may directly influence $Ca^{2+}$ oscillation system. When pancreatic β-cell is getting old, degradation of the function of mitochondria, SERCA and InsP3R will appear. According to the study of Sung-Hee Ihm et al in 2006[16], mitochondria ATP synthesis in β-cells will decline with age. And as Pall M L studied in 1990, not only ATP, but also ATP/ADP ratio in living cells will decline with age [17]. Therefore, with the growing of age, the level of ATP and available energy of ATP hydrolysis would both decline. Meanwhile, according to CJ Barker and et al's work in 2015[18], aging of β-cells will also slow down the rate of SERCA's transportation of $Ca^{2+}$. And InsP3R, as an important $Ca^{2+}$ channel for $Ca^{2+}$ oscillation, will also have some decline in its affinity to ligands [19].

According to the information above, we can do some stimulation of the effect of aging to $Ca^{2+}$ oscillation. After decreasing 15% of the rate constant of SERCA, decreasing 25% of the affinity of InsP3R to InsP3, and decreasing 25% of the rate constant of the dynamic of h, the oscillation area in ATP-ΔG graph is narrowed to where there the levels of ATP and ΔG are both relatively higher than before (Fig.8A). The amplitude and period of the oscillation (Fig.8B) will greatly declines, which fits the result of experiment by CJ Barker and et al pretty well.

Moreover, considering about the hypo-function of mitochondria, the level of ATP and ΔG will relatively decrease. According to the result shown in Fig.8A, the amplitude of oscillation will be lower than before or even stop oscillating.

# 4 Discussion

When ΔG is of a high level (~25RT), the regulatory effect of ATP is much stronger than ΔG. When its level is relatively low (~20RT), ΔG turns out to be an important regulatory factor.

How ΔG regulates the oscillation system is shown in two aspects. The first is that ΔG controls the amplitude of oscillation when the concentration of ATP is relatively low (Fig. 6). When ΔG grows higher, amplitude also becomes higher.

The second effect is high sensitive control in bifurcation. As is shown in Fig. 6, when ATP is relatively high, around the bifurcation point, where the level of ΔG is still low, bifurcation of this system shows high sensitivity to ΔG.

From those two conclusions above, we may draw the assumption that if a cell has problems in carbohydrate metabolism, the level of ΔG in it can be lower than usual, and those two phenomena discussed before may show in it. In this case, for $Ca^{2+}$ oscillation is a significant signal in living cells, a small change of ΔG may be vital to the whole cell.

This assumption can partly explain some experiment about the relationship between aging and the level of ATP and ATP/ADP ratio [17]. In fact, as mitochondria is an organelle who provides energy to the whole cell, the level of available energy should be more closed to the essence of how the level of ATP and ATP/ADP ratio controls the $Ca^{2+}$ oscillation in living cells.

And for the cells that are getting old, the function of SERCA, InsP3R and mitochondria will all be influenced. As is shown in Fig. 8, it would be much more difficult to generate the $Ca^{2+}$ oscillation in this kind of cells. In other words, these cells may be much more sensitive to the change of environment, such as the level of ΔG. This assumption may partly explain why diabetes is more likely occur among old people. As is mentioned before, in aged pancreatic β-cell, the level of ATP and ΔG is more closed to the boundary of oscillation area. When a slight perturbation of ATP or ΔG is added, which are most likely occur at the same time, aged β-cell is easier to get into the stable state, where the secretion of insulin will be inhibited.

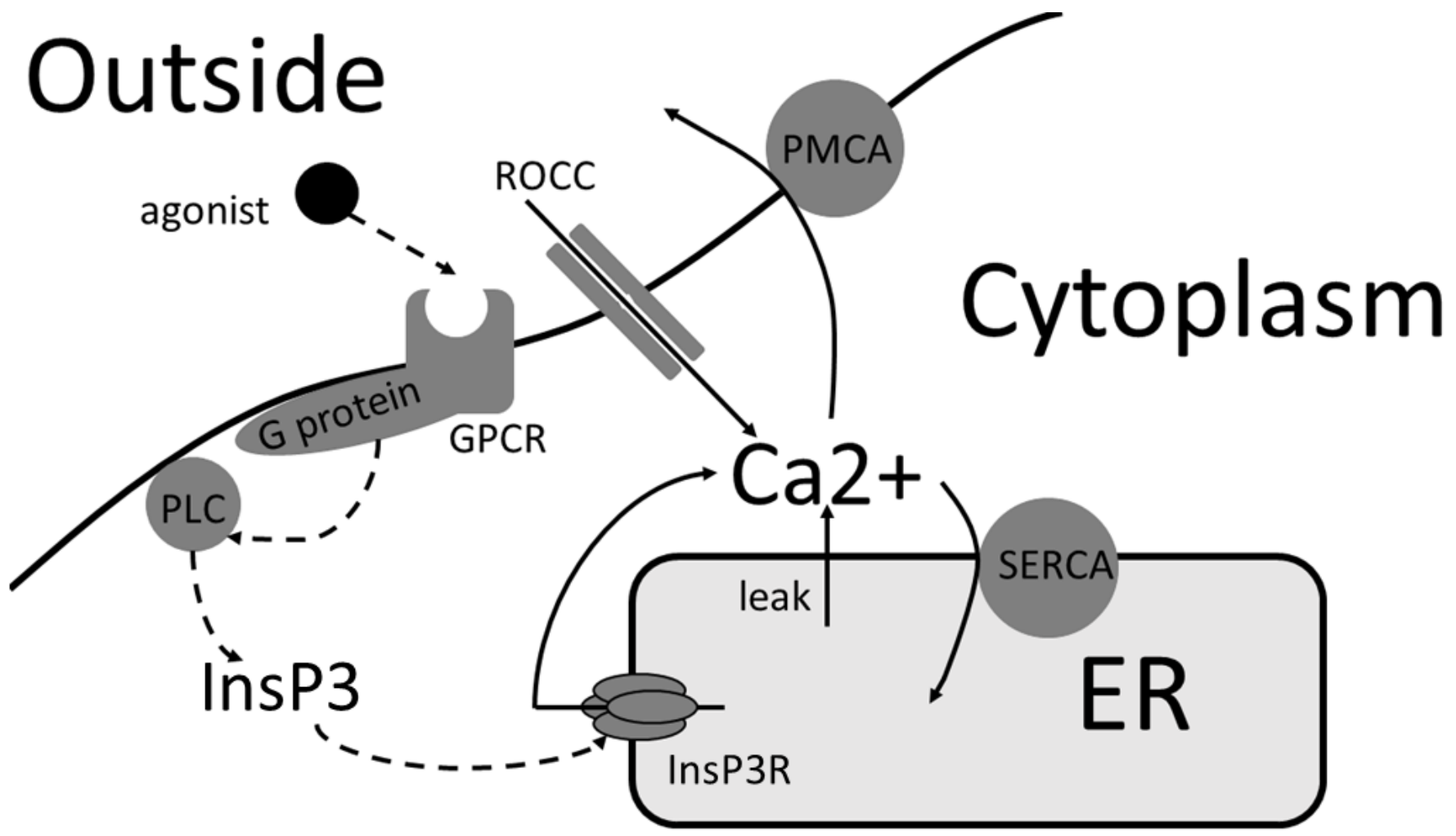

**Fig. 1 Total Model of Intracellular $Ca^{2+}$ Oscillation**. Binding of certain kinds of agonist activates both ROCC and GPCR. By a series of reaction, agonist will increase the concentration of InsP3 indirectly, which will manipulate the opening of InsP3R channel. Through ROCC, InsP3R and other secondary channels (written as leaking), $Ca^{2+}$ flows into cytoplasm. As the concentration of $Ca^{2+}$ increases, the function of SERCA/PMCA is enhanced until the flux through SERCA/PMCA become strong enough to decrease $Ca^{2+}$ concentration in cytoplasm to a lower level. When conditions are just right, oscillation will be stimulated.

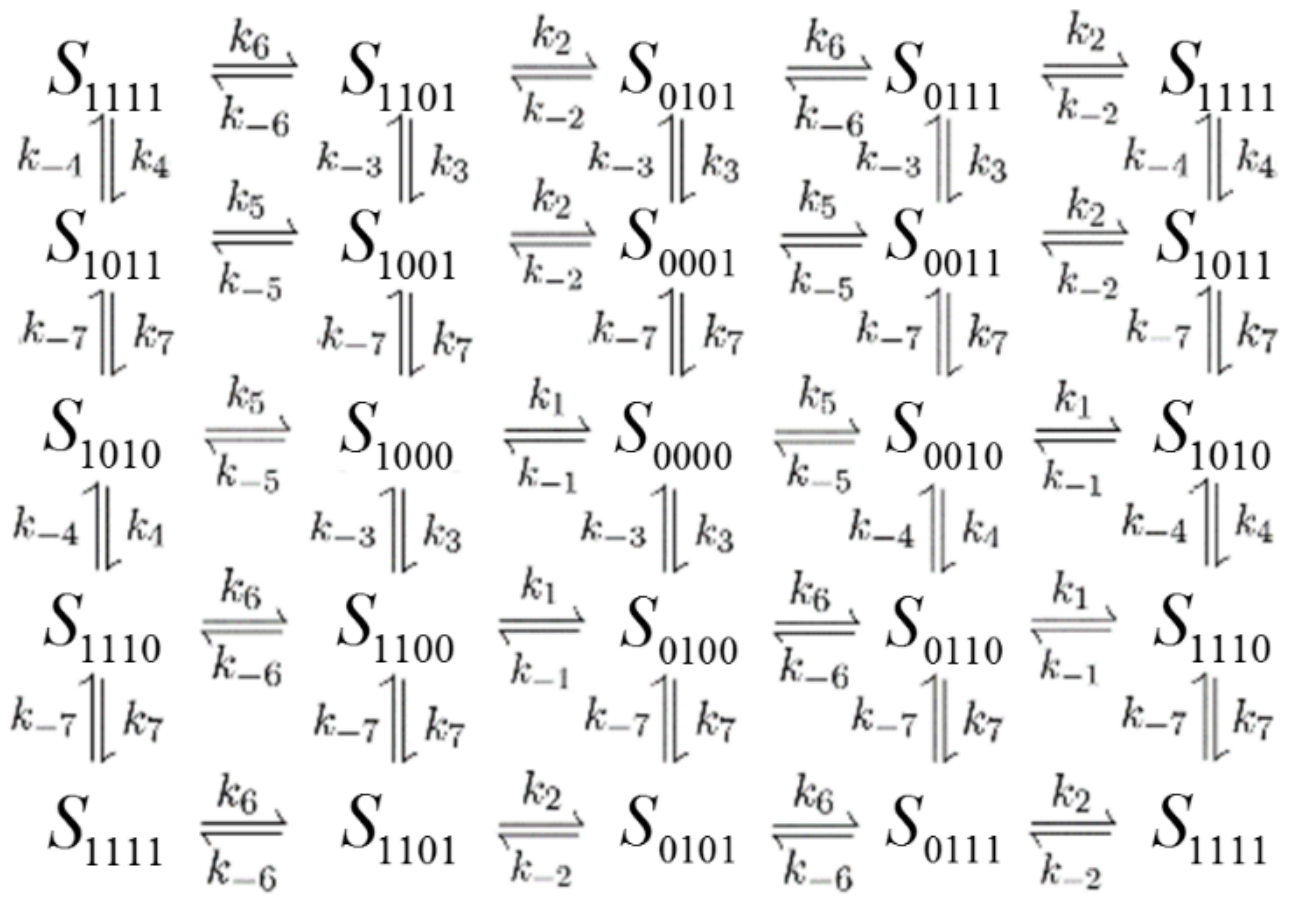

**Fig. 2 16-state model of InsP3 receptor.** The 4-dig footnote indicates the states of the four binding sites of InsP3R, where 1 is for captured state and 0 is for uncaptured state. And these four binary numbers indicates the states of $Ca^{2+}$ activation site, $Ca^{2+}$ inhibition site, InsP3 binding site, ATP binding site in sequence. Generally, when transitions happen at the same binding site in different states of InsP3R, they share the same reaction rate constants except in some certain cases. However, when ATP binding site is captured, the reaction rate of $Ca^{2+}$ activating site will change, and when InsP3 site is captured, the reaction rate of $Ca^{2+}$ inhibiting site will change.

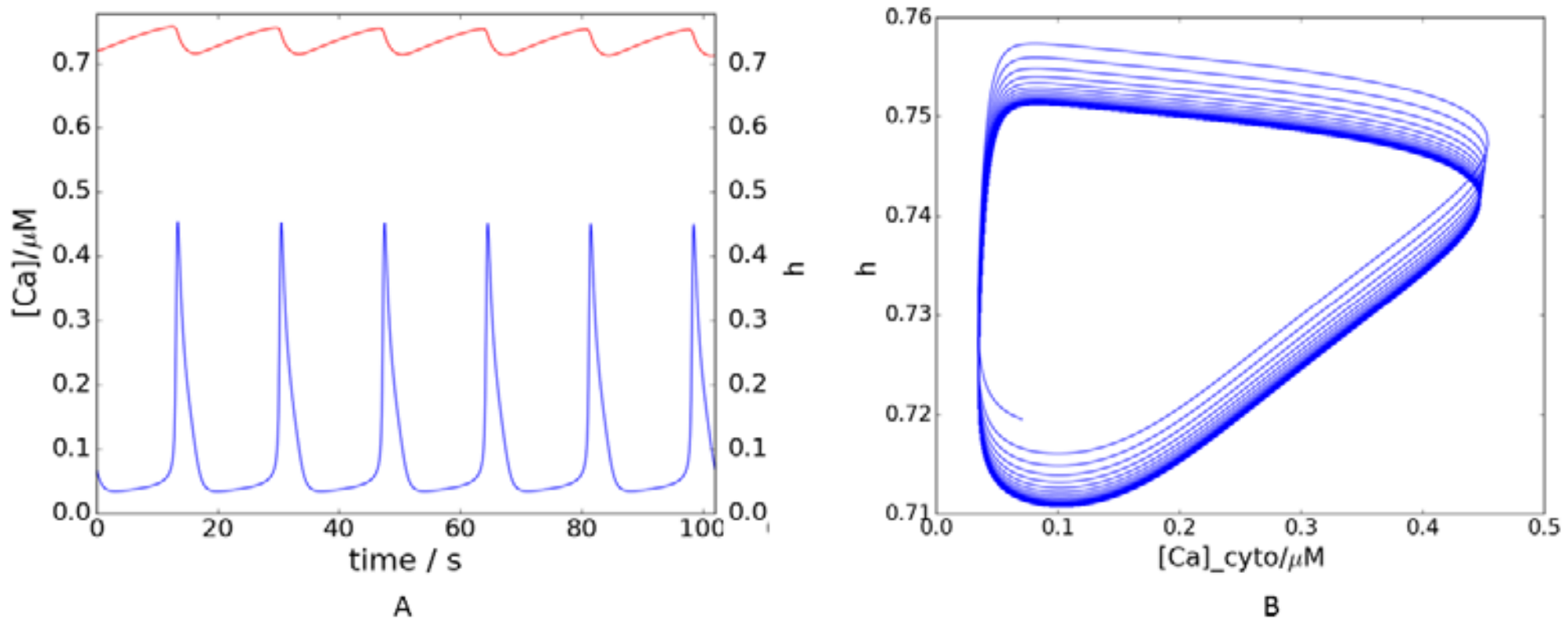

**Fig. 3** Graph A is the timing simulation of the oscillation system. The red line indicates the value of h, while the blue line indicates the concentration of $Ca^{2+}$ in cytoplasm. [ATP]=1600μM, ΔG=23RT and other parameters are shown in Table.1. Graph B is the corresponding phase portrait. Since it is a 3D system, null clines can't be shown in this graph. The period of the oscillation is around 20 seconds, and the peak of $Ca^{2+}$ reaches 0.4 μM, which has been confirmed by the wildly accepted experiment and simulation results.

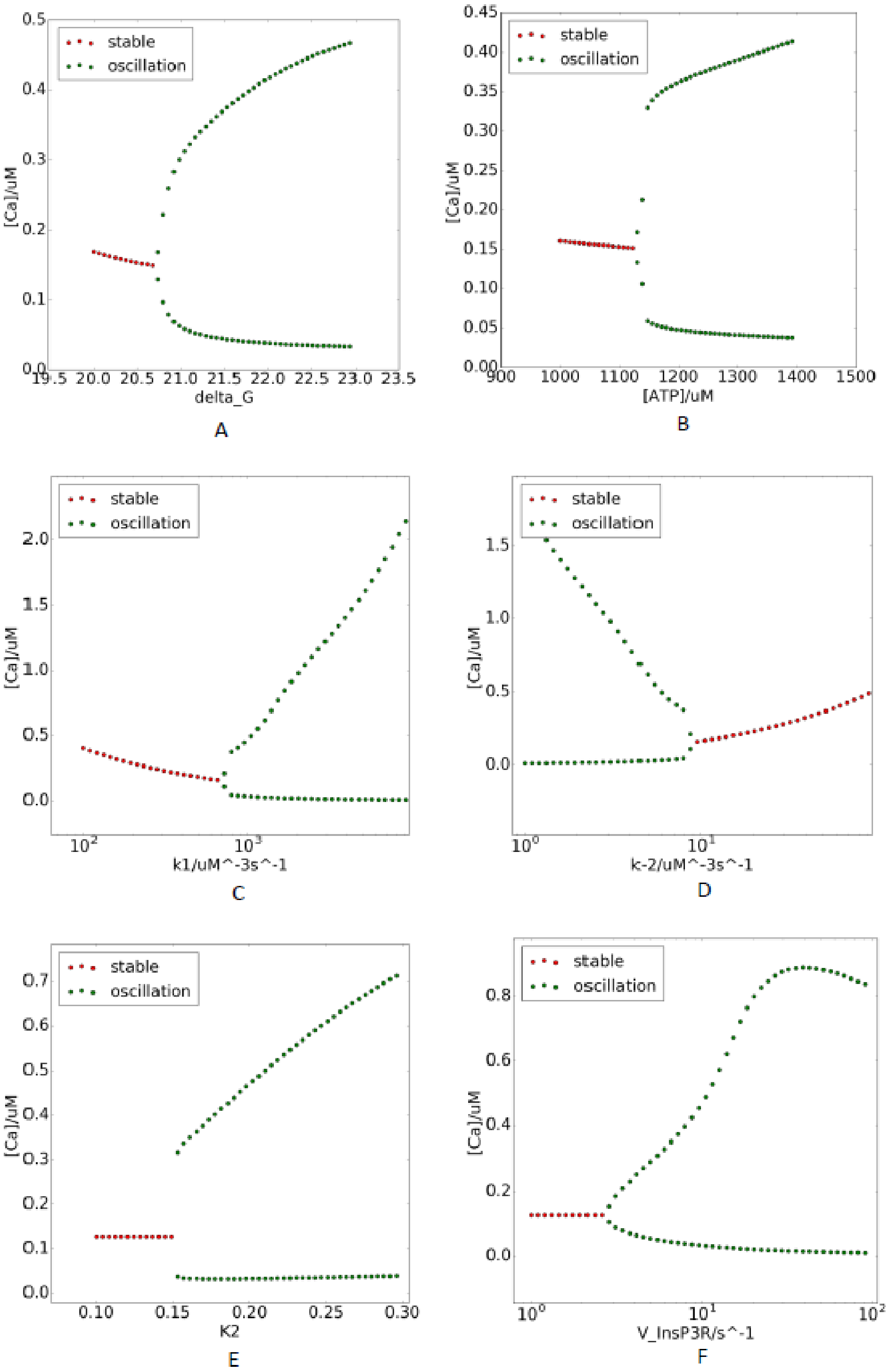

**Fig. 4 Bifurcation graphs of 6 sensitive parameters.** The names of parameters are shown in the labels of x axis. There exists Hopf bifurcations near the bifurcation points of these parameters. And as is shown above, ΔG and ATP concentration are much more sensitive than the other four, which means that these two parameters may play an important role in the manipulation of $Ca^{2+}$ oscillation.

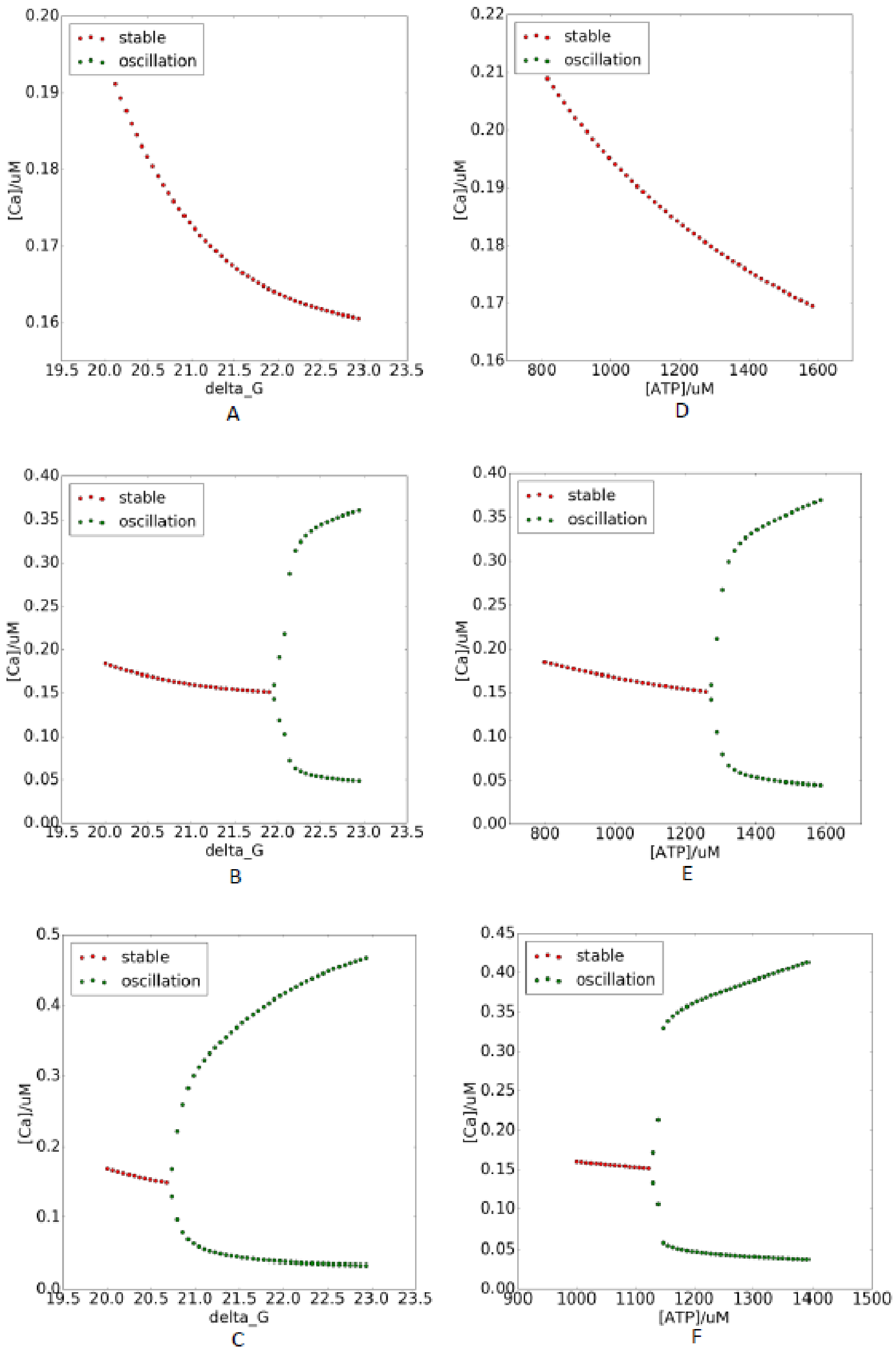

**Fig. 5 Bifurcation graphs of ΔG and ATP concentration.** Graph A, B, C show different bifurcation behavior of ΔG when ATP concentration changes, while graph D, E, F show the bifurcation behavior of ATP concentration with the change of ΔG. In graph A, B, C, ATP concentration equals to 1000μM, 1200μM and 1600μM; in graph D, E, F, ΔG equals to 20RT, 21.5RT, 23RT. Comparing graph A, B and C, with the growing of ATP concentration, the bifurcation point is moving to the left, similar situation happens when ΔG's value is growing. This means that ATP and ΔG have a strong correlation in the manipulation of $Ca^{2+}$ oscillation.

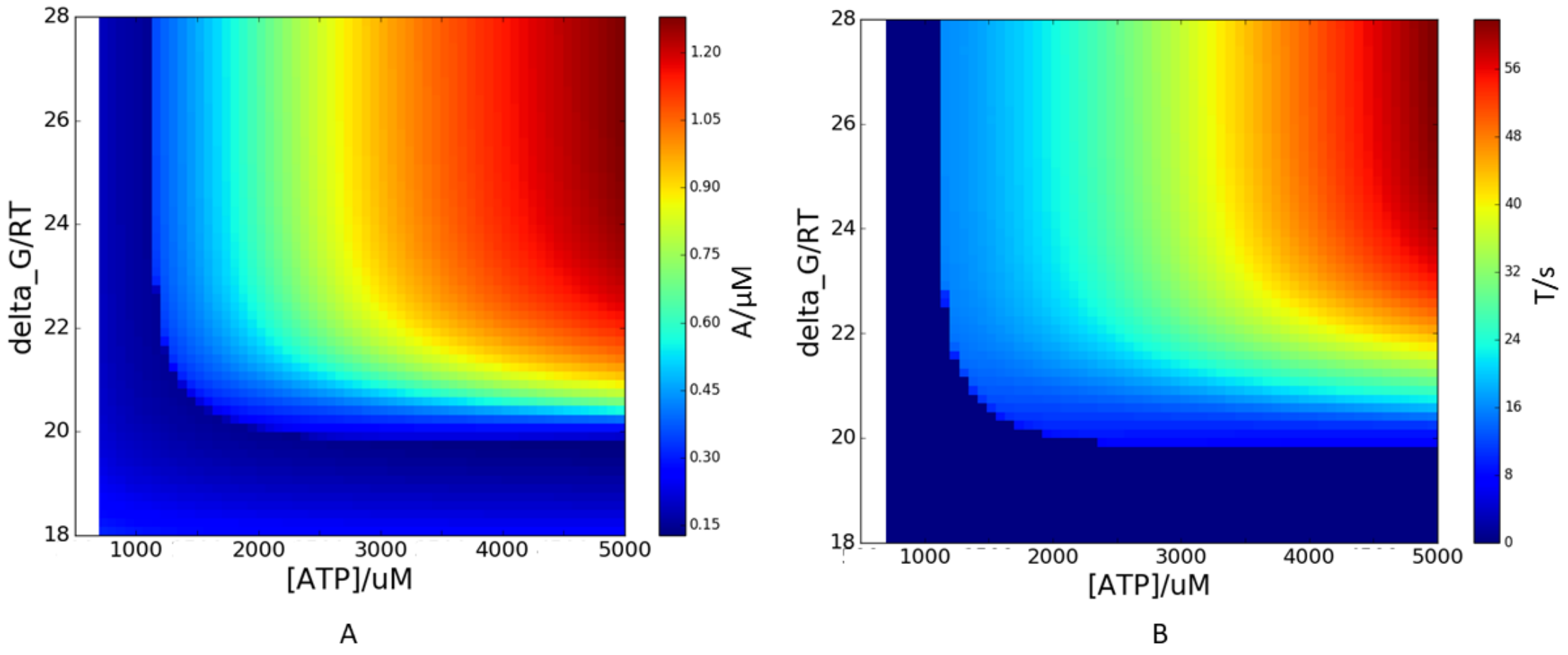

**Fig. 6 Oscillation amplitude (A) and period (B) influenced by ΔG and ATP concentration.** Correlation of ΔG and ATP concentration in $Ca^{2+}$ oscillation is shown more clearly in these two graphs. The value of color bars indicates the amplitude (A) and period (B) of $Ca^{2+}$ oscillation in cytoplasm. When any one parameter is near its bifurcation point, the amplitude and period of the oscillation will be markedly controlled by the other. And generally, amplitude and period of oscillation increases when ΔG and ATP concentration grows up. As it is shown in Fig. 5, the bifurcation behavior near the bifurcation point is hopf bifurcation. Meanwhile, there exists a minimum value of ATP and ΔG that the system can be in an oscillatory state only when their values are both high than the thresholds.

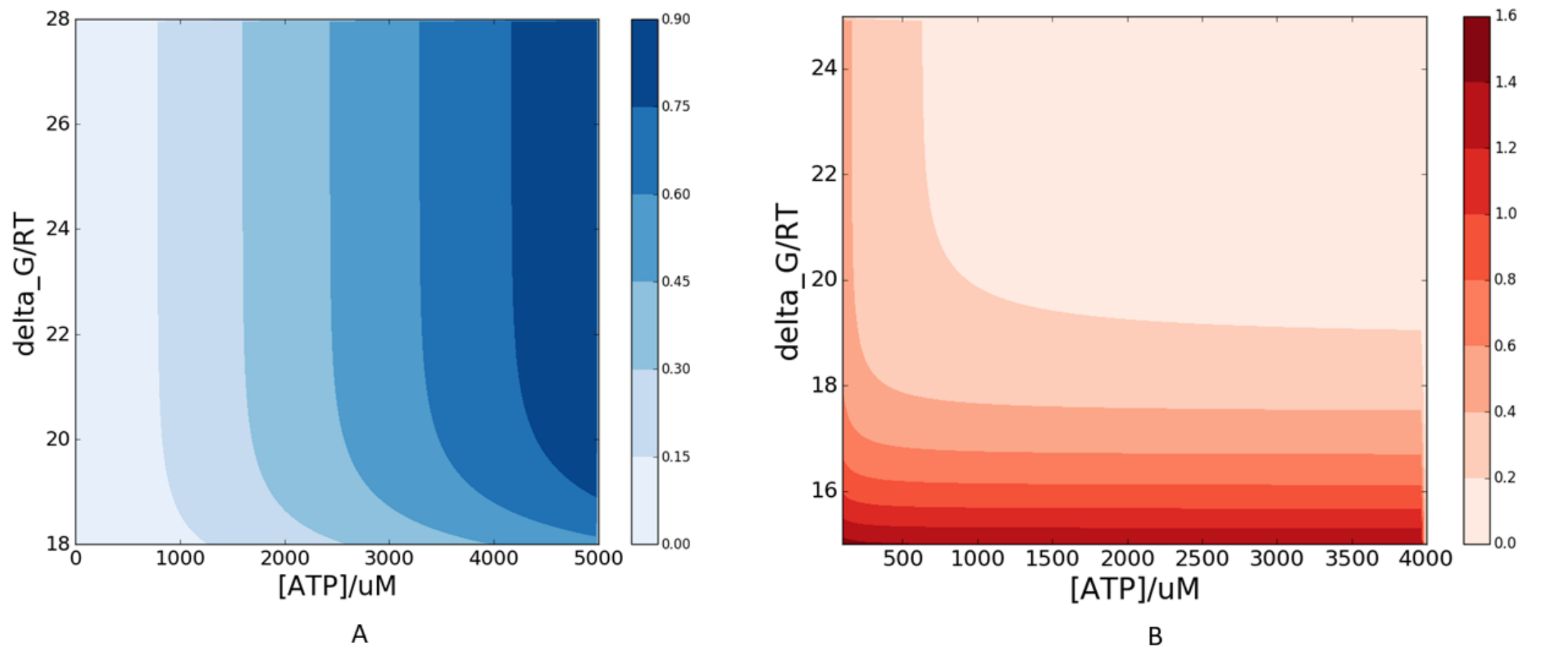

**Fig. 7 Analysis on the behavior of $Ca^{2+}$ pumps.** Graph A shows the change of the value of $I_{SERCA}$ with different [ATP] and ΔG, where $[Ca^{2+}{}_{cyto}]$=0.1μM and $[Ca^{2+}{}_{er}]$=10μM. Graph B shows stable $Ca^{2+}$ concentration in cytoplasm when IP3R is shut down. Since ATP hydrolysis only happens on $Ca^{2+}$ pumps (SERCA/PMCA), it makes sense to draw the conclusion that ΔG manipulates $Ca^{2+}$ oscillation by controlling the function of $Ca^{2+}$ pumps, which, however, may has little connection with ATP concentration. In this graph, the value of color bar indicates stable $Ca^{2+}$ concentration in cytoplasm.

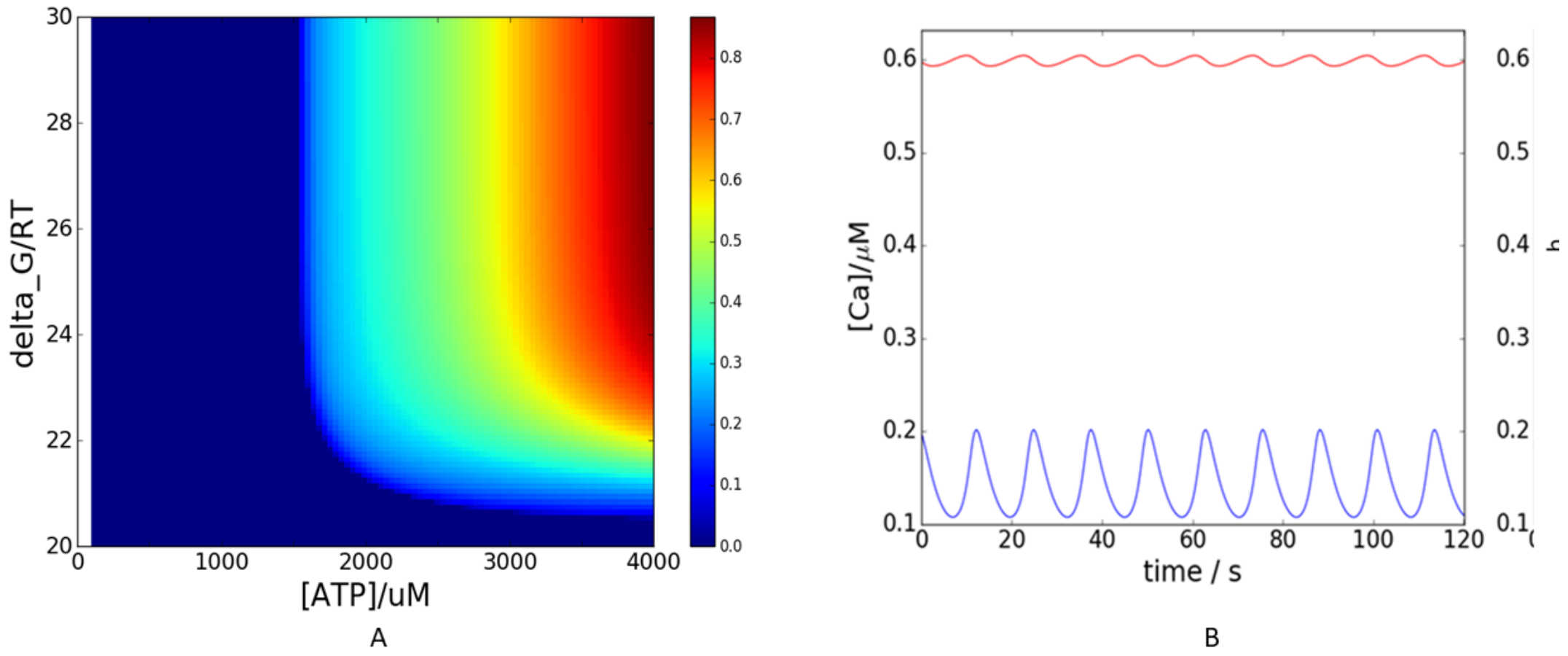

**Fig. 8 Simulation of $Ca^{2+}$ oscillation in aging cells.** These two graphs show how the system changes when the cell is getting old by decreasing 25% of the rate constant of SERCA, decreasing 25% of the affinity of InsP3R to InsP3, and decreasing 25% of the rate constant of the dynamic of h. In graph A, The area of oscillation is narrowed to a smaller space, where higher level of ATP and ΔG is needed to generate the oscillation. It is to say that for an old cell, it will be much more difficult to generate oscillation. In graph B, compared to Fig.3A, the amplitude of $Ca^{2+}$ oscillation in aging cells sharply decreases when the cell gets old, and the frequency, on the other hand, shows a slight increase.

# Acknowledgements

The authors are grateful to Liangyi Chen, Hong Qian, Huixia Ren and Chengsheng Han for helpful discussions.

# Appendix

## **A.** Mathematical derivation of 16-state model of InsP3R

According to a ligand-receptor model in equilibrium using partition function of thermodynamics [20], we can get the possibility of the existing of every single state when the InsP3R is in state h or (1-h), which contains 8 different states that are assumed in equilibrium state. First, the weight of each state could be represented by the reaction equilibrium constants and the concentration of corresponding ligands. In the state h, for example, we have:

| State | Weight |
|---|---|
| $\mathbf{S_{0000}}$ | 1 |
| $\mathbf{S_{0001}}$ | $[ATP]/K_7$ |
| $\mathbf{S_{0010}}$ | $[InsP3]/K_5$ |
| $\mathbf{S_{0011}}$ | $[ATP][InsP3]/K_7K_5$ |
| $\mathbf{S_{1000}}$ | $[Ca^{2+}]/K_1$ |
| $\mathbf{S_{1001}}$ | $[ATP]\ [Ca^{2+}]/K_2K_7$ |
| $\mathbf{S_{1010}}$ | $[InsP3]\ [Ca^{2+}]/K_1K_5$ |
| $\mathbf{S_{1011}}$ | $[InsP3]\ [ATP]\ [Ca^{2+}]/K_2K_7K_5$ |

Following the transition map of InsP3R (Fig.2), the kinetic function of h is:

$$\frac{dh}{dt} = -\phi_1 h + \phi_2 (1-h) \qquad \text{(A,1)}$$

$$\phi_1 = \frac{k_3[Ca^{2+}](S_{0000}+S_{0001}+S_{1000}+S_{1001}) + k_4[Ca^{2+}](S_{0010}+S_{0011}+S_{1010}+S_{1011})}{S_{0000}+S_{0001}+S_{0010}+S_{0011}+S_{1000}+S_{1001}+S_{1010}+S_{1011}}$$

$$= \frac{k_3[Ca^{2+}](1+[Ca^{2+}]/K_1+[ATP]/K_7+[Ca^{2+}][ATP]/(K_2K_7)) + k_4[Ca^{2+}][InsP3](1+[Ca^{2+}]/K_1+[ATP]/K_7+[Ca^{2+}][ATP]/(K_2K_7))/K_5}{(1+[Ca^{2+}]/K_1+[ATP]/K_7+[Ca^{2+}][ATP]/(K_2K_7))(1+[InsP3]/K_5)}$$

$$= \frac{k_3[Ca^{2+}] + k_4[Ca^{2+}][InsP3]/K_5}{1+[InsP3]/K_5}$$

$$\phi_2 = \frac{k_{-3}(S_{0100}+S_{0101}+S_{1100}+S_{1101}) + k_{-4}(S_{0110}+S_{0111}+S_{1110}+S_{1111})}{S_{0100}+S_{0101}+S_{0110}+S_{0111}+S_{1100}+S_{1101}+S_{1110}+S_{1111}}$$

$$= \frac{k_{-3}(1+[Ca^{2+}]/K_1+[ATP]/K_7+[Ca^{2+}][ATP]/(K_2K_7)) + k_{-4}[InsP3](1+[Ca^{2+}]/K_1+[ATP]/K_7+[Ca^{2+}][ATP]/(K_2K_7))/K_6}{(1+[Ca^{2+}]/K_1+[ATP]/K_7+[Ca^{2+}][ATP]/(K_2K_7))(1+[InsP3]/K_6)}$$

$$= \frac{k_{-3}[Ca^{2+}] + k_{-4}[Ca^{2+}][InsP3]/K_6}{1+[InsP3]/K_6}.$$

We assume that when monomer states $S_{1000}S_{1010}S_{1001}S_{1011}$ are activated, and only when four monomers are activated simultaneously, the calcium ions could go through InP3R. The open probability and transition intensity of the channel are

$$P_0 = (h\frac{S_{1000}+S_{1001}+S_{1010}+S_{1011}}{S_{0000}+S_{0001}+S_{0010}+S_{0011}+S_{1000}+S_{1001}+S_{1010}+S_{1011}})^4$$
$$= (h\frac{[Ca^{2+}]/K_1+[Ca^{2+}][ATP]/(K_2K_7)}{1+[Ca^{2+}]/K_1+[ATP]/K_7+[Ca^{2+}][ATP]/(K_2K_7)})^4 \qquad \text{(A,2)}$$

## **B.** Mathematical derivation of SERCA/PMCA

Taking SERCA as an example, $Ca^{2+}_{high}$ and $Ca^{2+}_{low}$ are replaced by $Ca^{2+}_{er}$ and $Ca^{2+}_{cyto}$, which respectively indicate $Ca^{2+}$ concentration in ER and cytoplasm. We have:

$$\frac{d[Ca^{2+}]_{cyto}}{dt} = 2k_{-1}E_2[ADP]-2k_1E_1[Ca^{2+}]^2{}_{cyto}[ATP] \qquad \text{(B,1)}$$
$$\frac{d[Ca^{2+}]_{er}}{dt} = 2k_2E_2-2k_{-2}E_1[Ca^{2+}]^2{}_{er}[Pi] \qquad \text{(B,2)}$$
$$\frac{dE_2}{dt} = -k_{-1}E_2[ADP]+k_1E_1[Ca^{2+}]^2{}_{cyto}[ATP]-k_2E_2+k_{-2}E_1[Ca^{2+}]^2{}_{er}[Pi] \qquad \text{(B,3)}$$
$$E_1+E_2=E_T \qquad \text{(B,4)}$$

Applying quasi-equilibrium assumption to the reaction of $E_2$, we have:

$$\frac{dE_2}{dt} = -k_{-1}E_2[ADP]+k_1E_1[Ca^{2+}]^2{}_{cyto}[ATP]-k_2E_2+k_{-2}E_1[Ca^{2+}]^2{}_{er}[Pi]=0 \qquad \text{(B,5)}$$
$$I_{SERCA}=\frac{d[Ca^{2+}]_{er}}{dt}=2E_T\frac{k_1k_2[ATP][Ca^{2+}]^2_{cyto}-k_{-1}k_{-2}[ADP][Pi][Ca^{2+}]^2_{er}}{k_1[ATP][Ca^{2+}]^2_{cyto}+k_{-2}[Pi][Ca^{2+}]^2_{er}+k_{-1}[ADP]+k_2} \qquad \text{(B,6)}$$

## **C.** Parameter sensitivity analysis

Sensitivity analysis of parameters of InsP3R , Ca2+-ATPase and ROCC will be shown in this part, We use the value in tables 1 to 4 as standard value, ATP=1600μM, and ΔG =23RT. To indicate the conditions of the oscillation system, we will compare the change of amplitude.

| PARAMETER | CHANGE | AMPLITUDE | CHANGE | AMPLITUDE |
|---|---|---|---|---|
| $\boldsymbol{k_1}$ | +10% | +11.68% | -10% | -10.91% |
| $\boldsymbol{k_{-1}}$ | +10% | -0.80% | -10% | +0.81% |
| $\boldsymbol{k_2}$ | +10% | +4.08% | -10% | -4.92% |
| $\boldsymbol{k}_{-2}$ | +10% | -9.93% | -10% | +13.03% |
| $E_{ER_T}$ | +10% | +1.14% | -10% | -1.24% |
| $E_{\boldsymbol{PM_T}}$ | +10% | +2.18% | -10% | -2.82% |
| $\boldsymbol{K_1}$ | +10% | +0.59% | -10% | -0.69% |
| $\boldsymbol{K_2}$ | +10% | +12.05% | -10% | -12.88% |
| $\boldsymbol{k_{+3}}$ | +10% | -2.02% | -10% | +2.07% |
| $\boldsymbol{k_{-3}}$ | +10% | +1.53% | -10% | -1.57% |
| $\boldsymbol{k_{+4}}$ | +10% | -6.43% | -10% | +6.92% |
| $\boldsymbol{k_{-4}}$ | +10% | +6.91% | -10% | -7.44% |
| $K_5$ | +10% | +2.23% | -10% | -2.40% |
| $K_6$ | +10% | -5.98% | -10% | +6.72% |
| $K_7$ | +10% | -0.03% | -10% | +0.03% |
| $V_{\boldsymbol{InsP3R}}$ | +10% | +8.34% | -10% | -8.17% |
| $[InsP3]$ | +10% | +3.88% | -10% | -4.13% |
| $V_{in}$ | +10% | -9.21% | -10% | +10.42% |
| $K_{in}$ | +10% | -0.02% | -10% | +0.02% |

Analysis shows that the system is not very sensitive to most parameters. Even for the most sensitive parameter, $\boldsymbol{K_2}$, when it is increased or decrease by 10%, the change of amplitude will be around only 12%. It means that the system is stable when parameters have some slight changes.